\documentclass[twocolumn]{aastex63}
\usepackage{amsmath}
\usepackage{graphicx}
\usepackage{graphicx,psfrag,epsfig,subfigure}
\usepackage{url}
\usepackage{indentfirst}
\usepackage{CJK}
\usepackage{txfonts}
\usepackage{paralist}
\usepackage{fancyhdr,lastpage}
\usepackage{color,hyperref}
\usepackage{pslatex}
\usepackage{amssymb}
\usepackage{afterpage}
\usepackage{upgreek}
\usepackage{epstopdf}

\makeatletter
\let\jnl@style=\rm
\def\ref@jnl#1{{\jnl@style#1}}
\makeatother

\newcommand*\patchAmsMathEnvironmentForLineno[1]{%
\expandafter\let\csname old#1\expandafter\endcsname\csname #1\endcsname
\expandafter\let\csname oldend#1\expandafter\endcsname\csname end#1\endcsname
\renewenvironment{#1}%
{\linenomath\csname old#1\endcsname}%
{\csname oldend#1\endcsname\endlinenomath}}%
\def\approxgt{\mathrel{\hbox{\rlap{\lower.55ex \hbox {$\sim$}}
        \kern-.3em \raise.4ex \hbox{$>$}}}}
\def\approxlt{\mathrel{\hbox{\rlap{\lower.55ex \hbox {$\sim$}}
        \kern-.3em \raise.4ex \hbox{$<$}}}}

\def\Msun{\hbox{$\rm ~M_{\odot}$}}

\def\H0{{\rm ~km~s^{-1}~Mpc^{-1}}}

\def\p9{{Pa$9$}}

\def\six{[\ion{S}{IX}]}

\def\L2-10{L$_{\rm 2-10keV}$}

\def\.25{0.25 keV\thinspace}

\def\d19{D$\,\leq\,$19~Mpc}

\newcommand{\swift}{\textit{Swift}}

\newcommand{\frb}{FRB\,121102}
\newcommand{\frbc}{\frb\ }

\submitjournal{APJ}

\begin{document}

\title{A search for short-term hard X-ray bursts in the direction of the repeating FRB\,121102}

\correspondingauthor{ Shangyu ~Sun and Wenfei~Yu}
\email{sysun@shao.ac.cn, wenfei@shao.ac.cn}

\author[0000-0002-0786-7307]{Shangyu ~Sun}
\affiliation{Key Laboratory for Research in Galaxies and Cosmology, Shanghai Astronomical Observatory, Chinese Academy of Sciences, \\
80 Nandan Road, Shanghai 200030, China.}

\author{Wenfei Yu}
\affiliation{Key Laboratory for Research in Galaxies and Cosmology, Shanghai Astronomical Observatory, Chinese Academy of Sciences, \\
80 Nandan Road, Shanghai 200030, China.}

\author{Yunwei Yu}
\affiliation{Institute of Astrophysics, Central China Normal University, Wuhan 430079, China}

\author{Dongming Mao}
\affiliation{Key Laboratory for Research in Galaxies and Cosmology, Shanghai Astronomical Observatory, Chinese Academy of Sciences, \\
80 Nandan Road, Shanghai 200030, China.}

\author{Jie Lin}
\affiliation{Key Laboratory for Research in Galaxies and Cosmology, Shanghai Astronomical Observatory, Chinese Academy of Sciences, \\
80 Nandan Road, Shanghai 200030, China.}

\begin{abstract}

The nature of fast radio bursts (FRBs), which occurs on millisecond time scales in the radio band, has not been well-understood. Among their unknown observational properties are their broadband spectra and persistent and transient multi-wavelength counterparts. Well-localized FRBs provide us the opportunity to address these issues in archival observations.  We have performed searches for 15-150 keV hard X-ray bursts on time scales as short as ${10}$\,milliseconds in the direction of the repeating FRB\,121102 (with a spacial resolution of a few arcminutes) in the archival \swift/BAT data during the period between October 2016 and September 2017.  We have found no significant (5\,$\sigma$) hard X-ray bursts in the direction of the repeating FRB.  We have derived an upper limit of the hard X-ray (15--150\,keV) flux of any X-ray bursts on 1 ms time scale of around $1.01 \times 10^{-7}$erg\,cm$^{-2}$s$^{-1}$, if assuming a photo-index of 2 for potential X-ray flares in X-ray band.  A plausible scenario for the repeating FRB as being associated with \emph{magnetar giant flare} is still far below the upper limit.
\end{abstract}


\section{Introduction} \label{intro}

Fast radio bursts (FRBs) are transient events of bright ($\sim$Jy) coherent radio emission on timescales as short as milliseconds \citep{2007Sci...318..777L,2013Sci...341...53T}. They are characterized by high dispersion measure (DM), which implies sources external to our Galaxy and likely at cosmological distances \citep{2003ApJ...598L..79I,2004MNRAS.348..999I}. Recent localization of several FRBs further strengthen their cosmological origin \citep{2019arXiv190611476B,2019arXiv190701542R}. FRBs are also characterized by a high event rate \citep[e.g. 2100\,day$^{-1}$ for GHz fluences $>2$\,Jy\,ms, estimated by ][]{2014ApJ...792..135T}. Although the event rate is high, up to now only tens detections of FRBs are known to the public, although the actual detections could be hundreds.

The FRB\,121102 had been the only known repeating FRB source \citep{2014ApJ...790..101S,2015MNRAS.454..457P,2016Natur.531..202S,2016ApJ...833..177S} before the detection of the second repeating FRB by CHIME \citep{2019arXiv190104525T}. The repetition of its bursts allowed a precise interferometric localization through the Karl G. Jansky Very Large Array (VLA) and the 350-m William E. Gordon Telescope at the Arecibo Observatory: right ascension (RA) 05\,hr\,31\,min\,58.70\,sec, declination (DEC) +33$^{\circ}$08$^{\prime}$52.5$^{\prime\prime}$ with a 1$\sigma$ uncertainty of about 0.1$^{\prime\prime}$, which is within 0.1$^{\prime\prime}$ of a faint 0.18-mJy persistent radio source with one-day-scale variability of 10\,$\%$ and angular size of $<1.7\times 10^{-3}$\,arcsecond \citep{2017Natur.541...58C}. Its DM in each burst reported in the studies of \cite{2016Natur.531..202S,2017Natur.541...58C} is consistent with the value of 558.1$\pm3.3$\,pc\,cm$^{-3}$. The DM of this source is three times the Galactic
maximum that has been predicted by the NE2001 electron-density model \citep{2002astro.ph..7156C}, and thus against a Galactic origin for the source. Assuming the Milky Way halo contribution to it DM$_{\rm halo}\approx 30$\,pc\,$^{-3}$, the disk contribution DM$_{\rm disk}\approx 188$\,pc\,$^{-3}$ \citep{2002astro.ph..7156C}, and the minimum of the host galaxy contribution DM$_{\rm host}= 0$, an upper bound for the source redshift can be obtained: $z\lesssim 0.32$, where the uncertainty in the mapping DM to redshift \citep{2014ApJ...780L..33M} is about 0.1. \citep{2017ApJ...834L...7T} have identified the host galaxy of \frbc with a redshift $z=0.19273$, consistent with the upper bound inferred from DM measurements. The VLA spectrum of the persistent radio counterpart is non-thermal and inconsistent with a single power-law \citep{2017Natur.541...58C}.

The millisecond durations and the huge energy releases of FRBs make them very likely related to the violent activities \citep[e.g., magnetar flares, giant pulses, and asteroid accretions, see ][]{2014MNRAS.442L...9L, 2016MNRAS.458L..19C, 2016ApJ...829...27D} or even global catastrophic collapses or mergers \citep{2014A&A...562A.137F,2014ApJ...780L..21Z,2018ApJ...868L...4M,2013ApJ...776L..39K,2013PASJ...65L..12T,2016ApJ...822L...7W,2016ApJ...826...82L, 2016ApJ...827L..31Z} of compact objects. Nevertheless, the repetition of \frbc undoubtedly enhances the connection of this FRB with a young active neutron star. The age of the neutron star can be constrained to be around one hundred years old, by its DM variation \citep[i.e., 10\% decreasing in seven months][]{2018Natur.553..182M} and the associated persistent radio emission \citep{2017ApJ...841...14M, 2017ApJ...839L..20C, 2017ApJ...838L...7D}.
Furthermore, a large rotation measures (RM) have been observed in \frbc \citep[$>10^{5}$\,rad\,m$^{-2}$; ][]{2018Natur.553..182M}, implying that this burst wave had travelled through highly magnetized plasma near the origin. The inferred strong magnetic field is much stronger than that in interstellar medium or intergalactic medium but could be consistent with a magnetar. Such a large RM had indeed been detected from the outburst of Galactic magnetars \citep[for e.g. SGR~J1745-2900][]{2013Natur.501..391E}.
The host galaxy of \frbc is a compact (diameter $\approxlt$ 4\,kpc) dwarf ($\sim 6 \times 10^{7}$\,\Msun) galaxy, which is very similar to those galaxies hosting superluminous supernovae or long-duration gamma-ray bursts \citep{2017ApJ...834L...7T}. A millisecond magnetar has been widely suggested to originate from these explosive phenomena, indicating them potential progenitors of repeating FRBs. Then an interesting suspect arises as that the FRBs might be associated by hard X-ray emissions that could also be driven by the young active magnetars.

Not only the nature of the radio bursts but also that of its potential persistent counterpart, has not been determined in any wavelength.  Apart from the search for potential persistent counterparts,  short-term transient counterparts for individual radio bursts in the time domain in wide range of wavelengths are valuable, such as searches in optical \citep{2017MNRAS.472.2800H} and ultra-high energies \citep{2018MNRAS.481.2479M}.  A systematic search for any short-term flaring events in the hard X-ray band ($>10$\,keV) which might be (or not) associated with the radio bursts with a random exposure to the repeating FRB\, 121102 is desired.  For the repeating FRB\,121102, the clustered occurrences of those radio bursts requires reasonable searches for any transient flares to spread in time. Two questions can be investigated. One is that whether there exists any hard X-ray bursts as short as several milliseconds coincident with the radio bursts. If not, what is the upper limit. The other question is that if we can find any association of hard X-ray bursts with radio bursts in terms of potential time lag between the events. This will help our understanding of the nature of the counterparts.

The \swift/BAT can cover the entire sky as efficiently as 80--90$\%$\,per day \citep{2013ApJS..209...14K} and offers a time resolution of $10^{-4}$ second for trigger event mode data. It is good for detecting or monitor short-and-bright X-ray transient sources, and therefore meets the requirement for addressing the aforementioned scientific questions.
As an initial effort to search for hard X-ray events associated with the majority of non-repeating FRBs, we present the method and our results achieved in our search for burst signals in the direction of FRB 121102 with the Swift/BAT data in Sect. 2. In order to search simultaneous X-ray signals, our search time interval was based on the burst time reports of FRB121102 that we had collected. We noticed there were burst reports in papers of \cite{2018Natur.553..182M} and \cite{2018ApJ...863....2G}: bursts from 2016-12-25 to 2017-08-26. Therefore, we select a search time interval containing these dates with burst reports. The total length of this search time interval is, however, limited by our computational algorithm for short-duration signal search. To search in Swift/BAT data of one year already cost us much time with our current algorithm. Base on these two reasons, we conduct our search in the period of October 1st, 2016 to September 30th, 2017 for this study.
Then we discuss and conclude in Sect.~\ref{dc}.

\section{Observations and analysis} \label{obser}

\begin{table}
\caption{Radio burst detected from \frbc during the 1-year period}\label{T:bursts}
\centering
\begin{tabular}{cccc}
\hline\hline
 & MJD$^{\diamond}$ & Observatory$^{\spadesuit}$ & $S^{\clubsuit}$[Jy] \\
\hline
1   &  57747.1295649013  &   Arecibo   & 0.9 \\
2   &  57747.1371866766  &   Arecibo   & 0.3 \\
3   &  57747.1462710273  &   Arecibo   & 0.8 \\
4   &  57747.1515739398  &   Arecibo   & 0.2 \\
5   &  57747.1544674919  &   Arecibo   & 0.2 \\
6   &  57747.1602892954  &   Arecibo   & 1.8 \\
7   &  57747.1603436945  &   Arecibo   & 0.6 \\
8   &  57747.1658277033  &   Arecibo   & 0.4 \\
9   &  57747.1663749941  &   Arecibo   & 0.2 \\
10  &  57747.1759674338  &   Arecibo   & 0.2 \\
11  &  57748.1256436428  &   Arecibo   & 0.1 \\
12  &  57748.1535244366  &   Arecibo   & 0.4 \\
13  &  57748.1552149312  &   Arecibo   & 0.8 \\
14  &  57748.1576076618  &   Arecibo   & 1.2 \\
15  &  57748.1756968287  &   Arecibo   & 0.4 \\
16  &  57772.1290302972  &   Arecibo   & 0.8 \\
17  &  57991.409904044   &    GBT      & 0.4  \\   
18  &  57991.412764720   &    GBT      & 0.05 \\    
19  &  57991.413019871   &    GBT      & 0.09 \\    
20  &  57991.413458764   &    GBT      & 0.3  \\   
21  &  57991.413706653   &    GBT      & 0.1  \\   
22  &  57991.413837058   &    GBT      & 0.2  \\   
23  &  57991.416436793   &    GBT      & 0.05 \\    
24  &  57991.416633362   &    GBT      & 0.7  \\   
25  &  57991.417714722   &    GBT      & 0.1  \\   
26  &  57991.417865553   &    GBT      & 0.1  \\   
27  &  57991.418627200   &    GBT      & 0.1  \\   
28  &  57991.419449885   &    GBT      & - -  \\   %
29  &  57991.421212904   &    GBT      & 0.1  \\    
30  &  57991.421712667   &    GBT      & 0.3  \\   
31  &  57991.422939456   &    GBT      & 0.1  \\   
32  &  57991.424270656   &    GBT      & - -  \\   %
33  &  57991.426552515   &    GBT      & 0.4  \\   
34  &  57991.430427904   &    GBT      & - -  \\   %
35  &  57991.431974007   &    GBT      & 0.1  \\   
36  &  57991.439360677   &    GBT      & 0.3  \\   
37  &  57991.448427650   &    GBT      & 0.1  \\   
\hline
\multicolumn{4}{l}{$^{\diamond}$ Modified Julian dates, which are referenced to infinite frequency at } \\
\multicolumn{4}{l}{\quad the Solar System barycentre; their uncertainties are of the order of} \\
\multicolumn{4}{l}{\quad the burst widths. Widths have uncertainties of about 10\,$\mu$s.} \\
\multicolumn{4}{l}{$^{\spadesuit}$ 110-m Robert C. Byrd Green Bank Telescope (GBT); } \\
\multicolumn{4}{l}{305-m William E. Gordon Telescope at the Arecibo Observatory} \\%
\multicolumn{4}{l}{$^{\clubsuit}$ Peak flux density} \\
\multicolumn{4}{l}{- - signal-to-noise ratio too low for flux calibration} \\
\multicolumn{4}{l}{These observations were reported in the papers of} \\
\multicolumn{4}{l}{\quad \cite{2018Natur.553..182M} and \cite{2018ApJ...863....2G}} \\
\end{tabular}
\end{table}
There had been quite some detections of radio bursts from \frbc in the period between Oct. 1st, 2016 and Sep. 30, 2017.  The bursts are listed in Table~\ref{T:bursts} based on reports in the literature. Since the radio bursts in the repeating FRB have a timescale of millisecond in the radio band, to search for high-energy short-term bursts in association with the occurrence of those radio bursts either simultaneously or not simultaneously, we have to make use of event mode data in the \swift/BAT archive. These BAT triggered event mode data have a time-resolution of about 100 microseconds, and the time stamp of each photon detected by BAT was recorded.

Initially we intended to search for potential X-ray bursts occurring simultaneously with the fast radio bursts as listed in Table~\ref{T:bursts}.  Unfortunately, the \swift/BAT archive does not have event data coincided with the occurrence of the radio bursts, and only observations in the survey mode coincided with the occurrences of those fast radio bursts.  We then decided to search for any X-ray bursts at any time occurring in the direction of \frbc in the event data in the period between Oct. 1st, 2016 and Sep. 30, 2017, composed of a total of 4770 triggered observations up to a total exposure of about 10300 ks, of which 17\% (in exposure time) were targeted at GRBs or GRB-like events, or GRB follow-ups.

The event mode data were processed by the software package \textsc{HEASOFT} v6.19 following \emph{The SWIFT BAT Software Guide}.  The \textsc{batdetmask} tasks were used to produce the detector quality map from CALDB. The \textsc{batdetmask} tasks calculate the mask weighting for each event file. The \textsc{batbinevt} tasks were run again to produced the light curves (LCs) from event files corresponding to the specified direction of \frb ~(RA 82.9946$^\circ$, DEC 33.1479$^\circ$) in the following four different energy bands, namely, 15--30, 30--60, 60--150, and 15--150\,keV.  To generate the sky maps, the \textsc{batbinevt} tasks were run to convert those event lists to detector plane images (DPIs), and the \textsc{batfftimage} task was used to covert them into sky maps with photon counts or significance.  As a result, the total exposure time of the event mode data towards \frb  ~in the period from Oct. 1st, 2016 to Sep. 30, 2017 was about 161 \,ks. The FRBs have ms timescales, and it is natural to search their X-ray conterparts in about ms timescales, but not limited in ms-timescale. Our search for potential short-term bursts was conducted on four representing timescales, namely 1 ms, 10 ms, 100 ms, and 1000 ms. We decide our shortest timescale is 1 ms because small binning results in plenty of LC/image data and large uncertainty in flux, and hence demands more computing resource.

\begin{figure*}
  \centering
  \subfigure[example of cuts] {
    \includegraphics[width=16.0cm,angle=0]{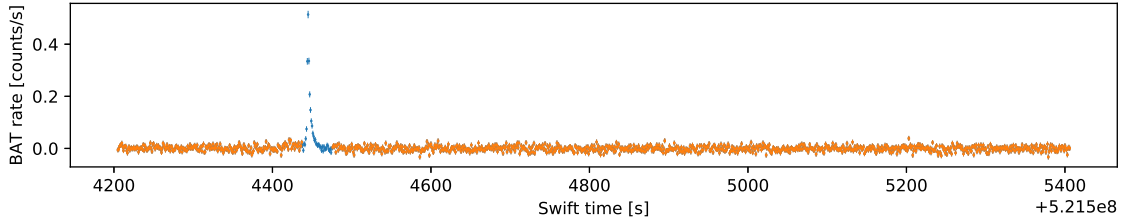}}

\caption{Here is an example showing the BAT event mode data corresponding to GRB 170711A used in our data analysis. Using the routine as described in the text, the segment of the data corresponding to the burst emission of GRB 170711A, shown in blue, was excluded. The data used in the search for any hard X-ray emission from the direction of FRB\,121102 is shown in orange.}
   \label{Fe00}
\end{figure*}




The event mode data are mostly taken due to on-board or uploaded triggers. However, not all the event data were used in our search for X-ray signals from the \frb,  because any bright sources or transients occurring in the field of view, e.g., gamma-ray bursts (GRBs), would affect our search for signals from the \frb. To improve the sensitivity for the detection of potential true X-ray flaring events from \frb ~ and to avoid false detections due to, e.g., the occurrence of gamma-ray bursts in the data, we have excluded the data corresponding to GRBs (within the T90 range of each GRB) from event data (see Fig.\,\ref{Fe00}, ). The T90 is the time over which the GRB emits from 5\% of its total measured counts to 95\%.
In addition, to exclude data when there are bright sources entering into the field-of-view of BAT,  the following segments were excluded from the original BAT event data in our search: (1) those segments corresponding to large fluctuations in the LCs (larger than 2 times the LC standard deviation $\sigma$ of the corresponded observation) at the beginning or at the end of the event data (within 5 seconds from the end), and (2) large dips occasional appears in the LCs with an interval within 5 seconds and an amplitude larger than 2 times the standard deviation. After applying these two conditions, about 85\% of the data remains.

Using the criteria above, we obtained the actual event data segments for our search. We performed a search for potential X-ray bursts in the resulted data set on 10 ms, 100 ms and 1000 ms time scales, respectively. The burst fluence is defined as the integral of the count rate in the LCs over a time range in which the count rates of all the time bins are above 0. Our search starts from the first time bin of each segments. Whenever the count rate of a certain bin is larger than zero, we then compute the product of this count rate and the time bin width. If the next  time bin has a count rate above zero, then we sum up the previous product and that of the current time bin. The summation continues until the count rate in the next time bin is no more than zero, and the resulted sum, the fluence of a potential burst, is recorded for further checking. Another summation starts again as we move on to the following time bins and find any bin has its rate above zero. This procedure goes on until every bin of this LC segment is checked.

After the computation of fluences of candidate bursts, we then check those bursts with fluences above a threshold. In this study, we set this fluence threshold as: \\
$5\times$ the standard deviation ($\sigma$) of count rates in the LC segment in examination $\times$ the time bin width. \\
Whenever there is a candidate burst with its fluence above the threshold, a sky map corresponding to the specific time range is then generated for checking whether the candidate burst has an astrophysical origin. In some cases, certain instrumental effects caused significant flux fluctuation at the edge of the sky map (due to small-portion illumination of the detectors by the sources at the edge of the field of view). In some cases, cosmic ray events probably caused wide-spread illumination of the detectors. We exclude these events and remove the corresponding time intervals in the LCs. However, at the end of the search, we found no significant astrophysical signal from the direction of \frbc during the period between October 1, 2016 and September 30, 2017 with the corresponding fluence larger than five times the count rate level of the standard deviation.


Here below we put constraints on their peak fluxes. The count rate ($r_i$) in the LCs were collected into histograms of distribution of count rates. We observed an non-Gaussian distribution, which indicates certain unknown systematic uncertainties.
In order to study statistical fluctuations alone so that we can put upper limits on the peak flux of potential bursts, we calculated the differential rate (i.e., $r_{i+1}-r_i$) instead to study statistical fluctuation. The differential rates in the period between October 1, 2016 and September 30, 2017 on time scale of 1 second are collected into a histogram, and the 1-$\sigma$ of the differential rate distribution corresponds to 0.058 [cts/s], obtained through the calculation of the root-mean-square (rms).
This is $\sqrt{2}$ times the standard deviation of the count rate distribution itself due to differential process if systematic effect is excluded. The 3-$\sigma$ upper limit of the count rate we derived was 0.125 [cts/s], which is
equivalent to $1.88 \times  10^{-8}$erg\,cm$^{-2}$s$^{-1}$ in the entire energy band of 15--150 keV if we assume an energy spectrum with a photon index of 2.
The upper limits on the count rates for the 15--30, 30--60, and 60--150\,keV bands are 1.46, 1.33, 1.26 $\times 10^{-8}$erg\,cm$^{-2}$s$^{-1}$. Similarly, the upper limit on 0.1 (0.01, 0.001) second time scales in the 15–150 keV is 3.48 (6.15, 10.1) $\times 10^{-8}$erg\,cm$^{-2}$s$^{-1}$.

\begin{figure*}
  \centering
  \subfigure[GRB]{
    \includegraphics[width=7.8cm]{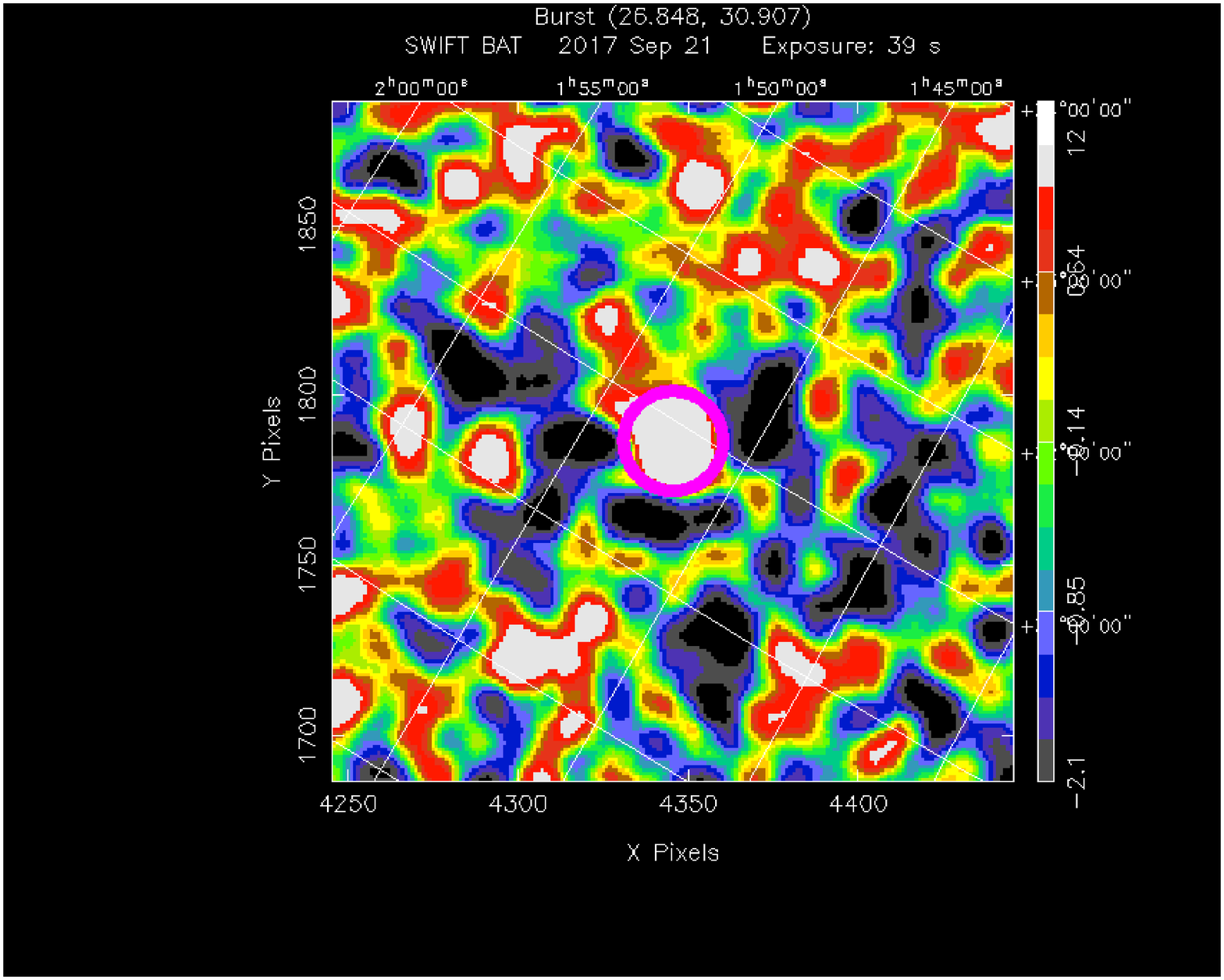}\label{Fe11a}}
  \subfigure[FRB]{
    \includegraphics[width=7.8cm]{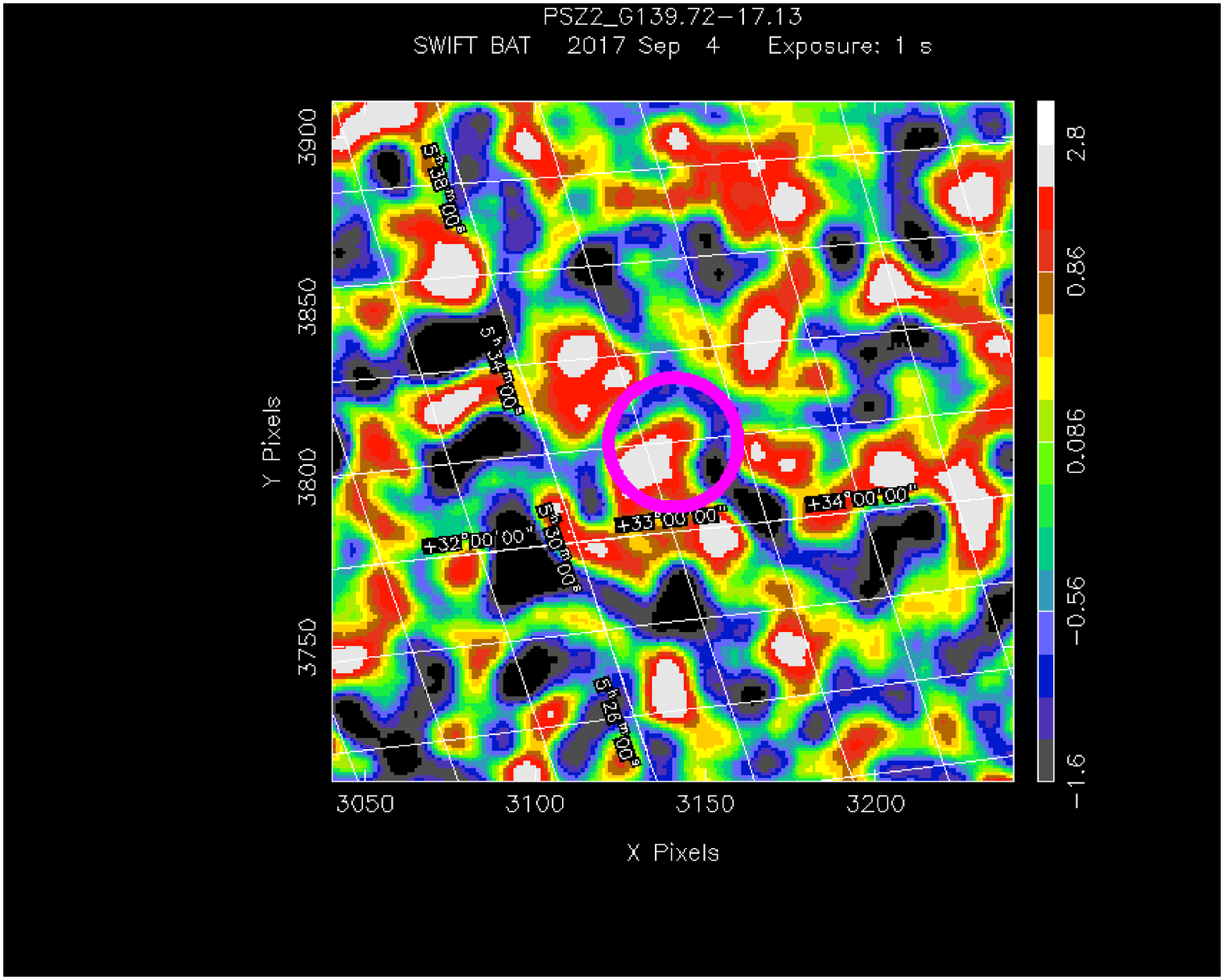}\label{Fe11b}}

    \caption{The BAT sky image during the GRB170921A and one of the best BAT image in the direction of \frbc taken over 1 second time scale. The pink circle marks the nominal size of the point spread function of BAT coded-mask imaging. }
   \label{Fe11}
\end{figure*}

\section{Discussion and conclusions} \label{dc}

We have searched for short-duration hard X-ray bursts in the direction of \frbc in the \swift/BAT archival event mode data. In the observations which covered the period from September 2016 to September 2017, neither a hard X-ray ($>$ 15 keV) burst occurring simultaneously to the reported radio bursts from the \frbc, nor a hard X-ray burst without an association of the reported radio bursts down to a time scale of 10 ms is detected in the direction of \frbc with a good spacial resolution of only a few arcminutes offered by \swift/BAT. We are able to put an upper limit of $1.01 \times 10^{-7}$erg\,cm$^{-2}$s$^{-1}$ of a hard X-ray burst in the energy band 15--150\,keV occurring in the direction of the repeating FRB on the time scale 1 millisecond, assuming a photo-index of 2. Slightly lower upper limits are obtained on 0.01, 0.1 and 1 second time scales.

In the period in which we searched for potential hard X-ray bursts associated with radio bursts, there have been 37 radio bursts detected as reported in the literature (see Table\,\ref{T:bursts}), and the actual number of FRBs occurred is therefore larger than 37 in a year's time. This yielded a lower limit of the radio burst event rate in \frbc, i.e., 37\,yr$^{-1}$. These 37 radio bursts were found in a total of 19.1 hours \frbc exposure time reported in relevant papers \cite{2018Natur.553..182M} and \cite{2018ApJ...863....2G}. The total Arecibo and GBT observation time for the search of bursts from FRB 121102 during October 1st, 2016 to September 30th, 2017 should be much longer than that reported in these two papers, given the fact that there should have been many observations performed without the detection of any radio bursts.  This leads us to derive an upper limit on the event rate as $1.7\times 10^{4}$ yr$^{-1}$. Base on the lower and upper limits of radio burst event rates, in the $\sim$161\,ks of effective BAT observations exposed towards \frb, the expected number of FRBs which should have been occurred is
between 0.19 and 87. Although these are rough estimates, the upper limits are more likely to relate to potential X-ray bursts simultaneous to the radio bursts.

We can compare these upper limits set by the one-year BAT observations in the event mode with those of theoretical predictions, in particular, with the model suggesting FRBs are accompanied by \emph{magnetar flares} \citep{2018ApJ...868L...4M}. The upper limit we derived on 1 second time scale obtained in the energy band 15--150\,keV is equivalent to $10^{-7}$erg\,cm$^{-2}$s$^{-1}$ if assuming a hard spectral index (say, $\alpha = 0$; since some quite extreme hard spectra in hard X-ray have been seen in magnetar flares). This means that the luminosity of the potential magnetar giant flare associated with \frbc cannot be higher than $\sim 10^{49}\rm erg\,s^{-1}$ on sub-second to second time scale for a luminosity distance of 970\,Mpc \citep{2017ApJ...834L...7T}. In other words, in such a framework, the ratio between the luminosities of the FRB radio burst and the X-ray giant flare should be higher than $\sim10^{-7}$ on second time scales. Therefore, if the previously observed Galactic magnetar X-ray giant flares can produce FRB emission in the same way, then the corresponding FRB radio luminosities could be as high as $>10^{37}-10^{40}$\,erg s$^{-1}$, corresponding to an average radio flux of about $0.1-100$ MJy for a distance of $\sim10$ kpc, in accordance with the observed X-ray flare luminosities ranging from $10^{44}$\,erg s$^{-1}$ to $10^{47}$\,erg s$^{-1}$ \citep{2017ARA&A..55..261K}. uch luminous FRBs, if indeed exist, are likely detected in our Galaxy in the recent past and the coming decades, since we have large field-of-view radio facilities running, such as the Australian SKA Pathfinder \citep[ASKAP; ][]{2008ExA....22..151J}, MeerKAT \citep[][]{2009IEEEP..97.1522J}, the Murchison Widefield Array \citep[MWA; ][]{2013PASA...30....7T}, the LOw-Frequency ARray \citep[LOFAR; ][]{2013A&A...556A...2V}, the Canadian Hydrogen Intensity Mapping Experiment \citep[CHIME; ][]{2018ApJ...863...48C}, and the next-generation Very Large Array \citep[ngVLA; ][]{2018ASPC..517....3M}. Several decades ago when those X-ray giant flares from SGRB 0526-66, 1900+14, and 1806-20 occurred, there were no wide field-of-view FRB radio facility operating. So potential association of the FRB bursts with magnetar giant flares can only be investigated in the current era when both wide field-of-view X-ray (such as Swift/BAT) and radio facilities (as mentioned above) are available. In the magnetar giant flare model, the FRB emission could
be produced by a synchrotron maser shock \citep{2014MNRAS.442L...9L,2017MNRAS.465L..30G,2017ApJ...843L..26B,2019MNRAS.485.4091M}, which is driven by the collision of a flare ejecta with a surrounding pulsar wind nebulae or with a following pulsar wind. As a further consequence, short-lived X-ray and gamma-ray afterglow emission are expected to be generated by the shock \citep{2019MNRAS.485.4091M}, the physics of which, in principle, can also be constrained by the upper limit of the X-ray peak flux. Nevertheless, at present, {\bf the upper limit we obtained is still at least $\sim10^4$ times higher than the predicted afterglow luminosity in the $10-150$ keV band that is around several times $10^{44}\rm erg~s^{-1}$ for typical model parameters (see Figure 8 in \cite{2019MNRAS.485.4091M})}. 
Finally, for a more general consideration, it is still necessary to investigate theoretically the possible high-energy emission in the other neutron star activity models (e,g., the giant pulse and the asteroid accretion models), which can make these models more verifiable.


\acknowledgments \label{Acknowledgments}

WY would like to thank Josh Grindlay of Harvard University and Craig Markwardt of GSFC for stimulating discussions in the past on Swift/BAT data. WY and SYS would like to thank David Palmer of LANL for very valuable comments on our previous analysis of BAT data. This work was supported in part by the National Program on Key Research and Development Project (Grant No. 2016YFA0400804) and the National Natural Science Foundation of China (grant number 11333005, U1838203, and 11822302). SYS also acknowledges support from China Postdoctoral International Recruitment Program in 2018. WY also acknowledges the support by the FAST Scholar fellowship, which is supported by special funding for advanced users, budgeted and administrated by Center for Astronomical Mega-Science, Chinese Academy of Sciences (CAMS). In this work, the authors used the data supplied by NASA's High Energy Astrophysics Science Archive Research Center in the US. The authors used the data products from the \swift mission, which is funded by NASA.


\bibliography{BibtexC}{}
\bibliographystyle{aasjournal}


\end{document}